# Realizing Bloch Dynamics in a Low-Cost Electrically Driven Acoustic Two-Level System

Xiao-Meng Zhang[1], Guang-Chen He[1], Zhao-Xian Chen[2,3], Ze-Guo Chen[1,3],*, Ming-Hui Lu[1,2,3],*, Yan-Feng Chen[2,3]

1School of Materials Science and Intelligent Engineering, Nanjing University, Suzhou 215163, China

2College of Engineering and Applied Sciences and Collaborative Innovation Center of Advanced Microstructures, Nanjing University, Nanjing 210093, China

3National Laboratory of Solid State Microstructures, Nanjing University, Nanjing 210093, China

*Corresponding author

E-mail addresses: zeguoc@nju.edu.cn (Ze-Guo Chen), luminghui@nju.edu.cn (Ming-Hui Lu)

**Abstract**

Unlike classical bits that can only occupy one of two discrete states, quantum bits (qubits) can exist in arbitrary coherent superpositions of the ground and excited states. This fundamental distinction grants qubits enhanced capabilities for information storage and processing. The Bloch sphere provides an intuitive and powerful geometric framework for visualizing, characterizing, and controlling the dynamical evolution of a qubit under external driving fields. By mapping the state evolution onto the Bloch sphere, processes such as spin flips and phase accumulation can be vividly represented as trajectories, enabling direct insight into coherent control mechanisms. Here, we implement Bloch dynamics in a classical platform by constructing a tunable acoustic two-level system based on high-quality-factor electro-acoustic coupled cavities. Using programmable spatiotemporal external field modulation, we demonstrate full Bloch sphere control through classical analogs of quantum phenomena, including Rabi oscillations, Floquet dynamics, Ramsey interference, and spin echo sequences. Our results bridge coherent Bloch dynamics with classical wave control, revealing a versatile experimental platform for exploring quantum-inspired physics. Furthermore, the system exhibits exceptional capabilities for precision transient acoustic field shaping, enabled by high-fidelity pulse-driven modulation.

# Introduction

While classical bits are limited to discrete binary states, quantum bits (qubits) can coherently exist in arbitrary superpositions of |0⟩ and |1⟩, offering a fundamentally richer state space for information processing[1]. Over the past decades, qubits have evolved from theoretical constructs into a variety of experimental platforms, including superconducting circuits based on Josephson junctions driven by microwave pulses[2–5], trapped-ion systems controlled by laser fields[6–10], and semiconductor quantum dots[11–13] manipulated using optical and microwave fields. The Bloch sphere serves as a complete and intuitive geometric representation of single-qubit pure states, providing a powerful framework for visualizing quantum superposition and for quantitatively describing coherent Bloch dynamics under external driving[14–20]. Core concepts of quantum control and Bloch dynamics have been innovatively extended into classical physical systems. For instance, coherent control in classical nanomechanical resonators has been demonstrated via electrostatic driving of orthogonal vibrational modes[21–23]; full Bloch-sphere-like control has been achieved in defect cavity pairs in photonic crystals by tuning the cavity separation[24]. However, both practical qubit control and classical two-level system (TLS) emulation are often constrained by demanding experimental conditions—such as strong coupling requirements, cryogenic environments, low quality factors, and limited tunability.

In contrast, engineered electro-acoustic cavities offer a promising alternative: they can support long coherence times under ambient conditions and eliminate the need for extreme setups[25–27]. When combined with programmable, time-varying control circuitry, these systems enable flexible and high-precision manipulation of TLS dynamics across the full Bloch sphere. Here, we experimentally realize a classical electro-acoustic two-level system and demonstrate its transient Bloch dynamics under externally programmable driving fields. By mapping Rabi oscillations, Floquet dynamics, and composite pulse sequences onto the Bloch sphere, we track the full state evolution and its transient response, achieving high-precision and tunable Bloch dynamics of transient acoustic field. This work bridges quantum-inspired Bloch sphere control frameworks with classical wave systems, paving the way for advanced acoustic technologies and interdisciplinary exploration of coherent dynamics.

**Results**

A key challenge in realizing Bloch dynamics lies in the precise and programmable manipulation of acoustic fields—far beyond conventional adjustments of intensity or frequency. It requires real-time shaping and control of the sound field with high precision and flexibility. To address this challenge, we engineer a classical acoustic two-level system, as illustrated in Fig. 1a. Two orthogonal resonant modes at frequencies $\omega_0$ constitute the basis states $|0\rangle = (1,0)^T$ and $|1\rangle = (0,1)^T$, respectively. The instantaneous acoustic wavefunction of the system can then be expressed as a coherent superposition of these basis states: $|\psi(t)\rangle = c_0(t)|0\rangle + c_1(t)|1\rangle$, with complex amplitudes $c_0(t) = \cos\left(\frac{\theta(t)}{2}\right), c_1(t) = e^{i\varphi(t)}\sin\left(\frac{\theta(t)}{2}\right)$. Thus, the dynamical evolution of the TLS state can be intuitively visualized as a trajectory on the Bloch sphere. Real-time monitoring of the mode amplitudes and their relative phases allows us to reconstruct and precisely manipulate the state trajectory, represented by the Bloch vector $(x, y, z) = (\sin\theta(t)\cos\varphi(t), \sin\theta(t)\sin\varphi(t), \cos\theta(t))$. A time-reconfigurable circuit network (as shown in Fig. 1b) synthesizes a two-level time-dependent Hamiltonian control the TLS dynamics:

$$H(t) = 2\pi \begin{pmatrix} \Delta(t)/2 - i\gamma_1(t) & \kappa(t) \\ \kappa(t) & -\Delta(t)/2 - i\gamma_2(t) \end{pmatrix} \qquad (1)$$

where the detuning $\Delta(t)$, cavity loss factor $\gamma_1(t), \gamma_2(t)$ and coupling strength $\kappa(t)$ are precisely modulated in time via an active electro-acoustic feedback architecture. This system integrates voltage-controlled amplifiers and digital phase shifters within real-time feedback loops, enabling high-fidelity control over acoustic field parameters through programmable electronic circuits (see Supplementary S1 for detailed implementation). To elucidate the transient state manipulation on the Bloch sphere, we begin our analysis by focusing on the lossless driving component:

$$H_{drive}(t) = 2\pi \left(\frac{\Delta(t)}{2}\hat{\sigma}_z + \kappa(t)\hat{\sigma}_x\right). \qquad (2)$$

The state evolution is governed by the time-dependent Schrödinger-type equation: $i\frac{d}{dt}|\psi(t)\rangle = H_{drive}(t)|\psi(t)\rangle$. By precisely modulating $H_{drive}(t)$, flexible driving schemes—such as continuous, composite, or periodic controls—can be implemented to achieve versatile state manipulation on the Bloch sphere. In particular, continuous driving with the Hamiltonian

$H_{drive} = 2\pi\kappa\sigma_x$ generates state rotations around $x$-axis. Precise control of driving duration $t = 1/(4\kappa)$ can realizes a complete population inversion from $|0\rangle$ to $|1\rangle$ ($\pi$ pulse), while a duration $t = 1/(8\kappa)$ yield a $\pi/2$ pulse, transforming $|0\rangle$ to $(|0\rangle - i|1\rangle)/\sqrt{2}$. Alternatively, introducing a detuning ($H_{drive} = \pi\Delta\hat{\sigma}_z$) with disabled coupling results in state rotations around the $z$-axis, enabling accurate control of the relative phase $\varphi(t)$; this process is commonly referred to as free precession (see Supplementary S2 for detailed derivations). More flexible and sophisticated control can be realized by sequentially switching and combining multiple $H_{drive}$ with distinct time-dependent profiles. Such composite driving protocols integrate rotations about multiple axes and controlled phase accumulation, enabling precise multi-degree-of-freedom modulation on millisecond timescales and facilitating accurate spatiotemporal tailoring of wavefunction evolution paths, even within classical acoustic contexts.

**Rabi oscillations and chiral evolution path**

As shown in Fig. 1c, we present both theoretical and experimental results for the time-dependent population evolution of state $|0\rangle$, driven by the $H_{drive} = 2\pi\left(\frac{\Delta}{2}\hat{\sigma}_z + \kappa\,\hat{\sigma}_x\right)$. The observed oscillations confirm the formation of a coupled classical two-level system. While conventional studies have primarily focused on population transfer in Rabi oscillations[28–31], the phase evolution and the symmetry properties of the corresponding Bloch sphere trajectories have received comparatively less attention. Here, we explicitly demonstrate how the externally controlled phase accumulation induces asymmetric, chirality-dependent evolution trajectories on the Bloch sphere, governed by the sign of detuning and the choice of initial state. Under driving Hamiltonian with opposite and fixed detuning, the wavefunctions evolve as: $|\psi_\pm(t)\rangle = \cos\left(\frac{\Omega t}{2}\right)|0\rangle \pm i e^{\pm i\Delta t}\sin\left(\frac{\Omega t}{2}\right)|1\rangle$ ($\Omega = \sqrt{(2\pi\kappa)^2 + (2\pi\Delta)^2}$). By extracting this phase information, we directly project the evolution trajectories onto the Bloch sphere, clearly illustrating asymmetric phase accumulation and the formation of chiral paths rotates around $\hat{n} = \frac{(2\kappa,0,\Delta)}{\sqrt{(2\kappa)^2+\Delta^2}}$ (Fig. 1d). Specifically, starting from $|0\rangle$ state, a positive detuning ($\Delta = +20$ Hz) induces clockwise precession around $\hat{n}$, corresponding to forward phase accumulation, whereas negative detuning ($\Delta = -20$ Hz) leads to a counterclockwise precession and reversed phase accumulation. The two trajectories are spatial mirror images of each other, but exhibit opposite chirality. Conversely, when starting from the initial state $|1\rangle$, the evolution trajectories under opposite detuning are also spatially mirrored, but relative to those originating from $|0\rangle$, they exhibit reversed rotation directions while preserving the same chirality structure.

Despite the precise programmability of pulse-driven control, two key relaxation processes—energy dissipation and phase decoherence—fundamentally limit system performance[22,32]. These mechanisms constrain the lifetime of TLS dynamics and set intrinsic limits on high-fidelity state manipulation. To prolong the energy relaxation time $T_1$ and suppress energy dissipation, we implement an active feedback circuit that enables dynamic control of the system's acoustic response. Notably, it enables signal amplification facilitating high-quality-factor resonances. As illustrated in Fig. 1e (S1–S4), initialization in the $|0\rangle$ state followed by synthetic-field-engineered $\pi$ pulse excitation demonstrates tunable modulation of $T_1$.

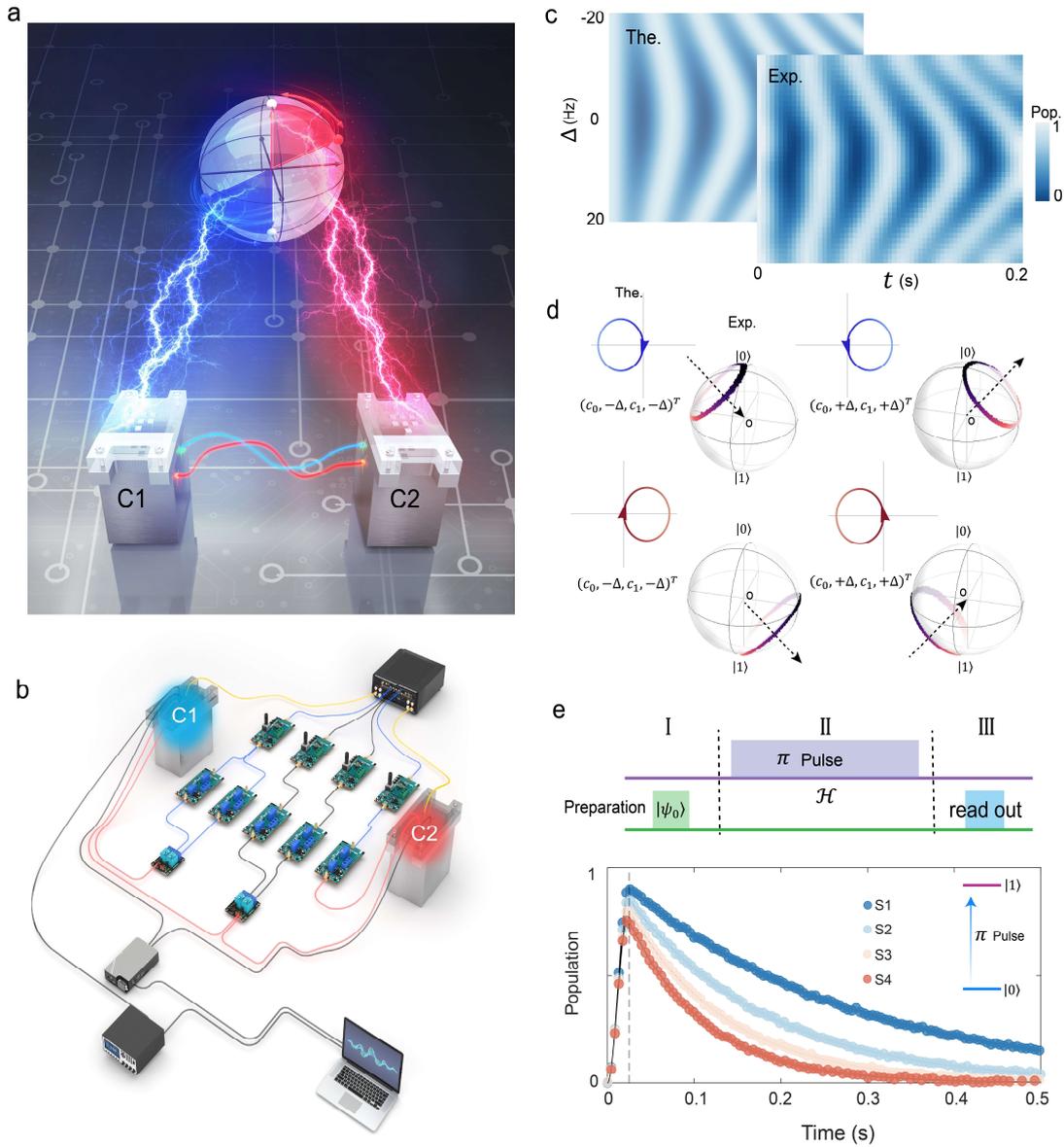

**Fig. 1| Classical electrical-acoustic two-level system demonstrating Bloch dynamics.**

**a** Schematic of a classical electro-acoustic coupled two-level system. **b** Experimental platform for state manipulation via external electric field modulation, featuring a high-fidelity circuit network and programmable voltage control units for encoded periodic voltage modulation. **c** Theoretical and experimental Rabi oscillations of the state $|0\rangle$ under varying detuning and evolution times ($\kappa = 10$). **d** Chiral evolution trajectories on the Bloch sphere: Top row–theoretical and experimental paths for the system initialized in $|0\rangle$ under positive/negative detuning, showing clockwise/counterclockwise evolution; Bottom row–corresponding trajectories for the $|1\rangle$ initial state, demonstrating detuning-dependent chirality reversal ($\kappa = 10$). **e** State preparation, pulsed excitation, and readout processes regulated by a positive-feedback circuit, yielding measured $T_1$ relaxation times (S1~S4: 0.53 s, 0.32 s, 0.22 s, 0.17 s).

**Floquet dynamics**

Beyond hardware-level feedback control, Hamiltonian engineering via Floquet periodic driving provides an alternative and powerful route to suppress or delay energy relaxation[33–36]. Building on this idea—and to access more complex dynamical regimes, we incorporate Floquet modulation into our experimental platform to explore its influence on coherent state evolution. The two resonant modes $|0\rangle$ and $|1\rangle$ are engineered with distinct $T_1$ relaxation times ($Ts_1$ and $Ts_2$, respectively, as schematically illustrated in Fig. 2a). By periodically inverting $Ts_1$ and $Ts_2$, the system Hamiltonian satisfies $H(t+T) = H(t)$, establishing a Floquet-driven scenario, and the Hamiltonian can be conveniently rewritten as:

$$H(t) = \begin{cases} H_1 = 2\pi \begin{pmatrix} i\delta\gamma & \kappa \\ \kappa & -i\delta\gamma \end{pmatrix} - i2\pi\gamma I, & \text{for } 0 \leq (t \bmod T) < \alpha T \\ H_2 = 2\pi \begin{pmatrix} -i\delta\gamma & \kappa \\ \kappa & i\delta\gamma \end{pmatrix} - i2\pi\gamma I, & \text{for } \alpha T \leq (t \bmod T) < T \end{cases} \quad (3)$$

Here we define $\gamma_{1,2} = \frac{1}{Ts_{1,2}}$, $\gamma = \frac{i(\gamma_1+\gamma_2)}{2}$, and $\delta\gamma = \frac{i(\gamma_1-\gamma_2)}{2}$. The parameter $\alpha \in (0,1)$ specifies the duty cycle of the Hamiltonian $H_1$ within each modulation period. In this work, we fix $\alpha = 0.5$. The time evolution of the wavefunction can be described as $|\psi(t)\rangle = e^{-i\varepsilon t}e^{-i2\pi\gamma}|\phi(t)\rangle$, where $\varepsilon$ represents the quasi-energy and $|\phi(t)\rangle$ is a periodic wavefunction with period $T$. Omitting a passive loss term $i2\pi\gamma I$, we consider a $\mathcal{PT}$-symmetric Hamiltonian component $H_{\text{pt}} = 2\pi \begin{pmatrix} \pm i\delta\gamma & \kappa \\ \kappa & \mp i\delta\gamma \end{pmatrix}$ which governs energy gain/loss balance and coherence lifetime. The Floquet quasi-energy spectrum is obtained by diagonalizing the one-period time-evolution operator $U = \mathcal{T}\exp\left(-i\int_0^T H_{\text{pt}}(t)dt\right)$, where $\mathcal{T}$ denotes the time-ordering. The eigenvalue equation $U(T)|\psi\rangle = e^{-i\varepsilon T}|\psi\rangle$ defines the quasi-energies $\varepsilon$. Under Floquet modulation, the exceptional point (EP) condition $\delta\gamma = \kappa$ is relaxed and replaced by a modified threshold[34]: $\cos^2\left(\frac{2\pi\sqrt{\kappa^2-(\delta\gamma)^2}T}{2}\right) = \frac{(\delta\gamma)^2}{\kappa^2}$ (see details in Supplementary S3). This result reveals that the Floquet EP depends jointly on $\kappa, \delta\gamma,$ and $T$, enabling effective gain at much lower $\delta\gamma$ than in the static case. As shown in Fig. 2b, for $\kappa = 8.5$ even small $\delta\gamma$ can induce imaginary quasi-eigenenergy (white regions) when $T$ is tuned appropriately. A clear example is given in Fig. 2c, where Floquet modulation with $T = 0.06$ s reduces the gain threshold to $\delta\gamma = 0.27$, compared to $\delta\gamma = 8.5$ in the unmodulated case. This substantial threshold reduction allows tiny variations in $\delta\gamma$ to trigger mode-selective amplification or decay, enabling net gain per cycle and compensating intrinsic losses. As a result, Floquet engineering extends state lifetimes and enhances signal fidelity, offering a powerful tool for robust Rabi oscillations and dynamic field control. This theoretical prediction is experimentally confirmed, as shown in Fig. 2d: the transient responses of the two coupled resonators (C1 and C2) exhibit sharply contrasting behaviors under static and Floquet-modulated conditions. Without modulation, both cavities display rapidly damped oscillations, whereas Floquet modulation sustains coherent oscillations over significantly longer timescales, indicating enhanced stability. Figure 2e quantitatively illustrates this effect via the temporal evolution of normalized acoustic power. Under Floquet modulation, the power decay is substantially slower than in the static case, revealing modulation-induced suppression of dissipation and the emergence of quasi-stationary states stabilized by the driving.

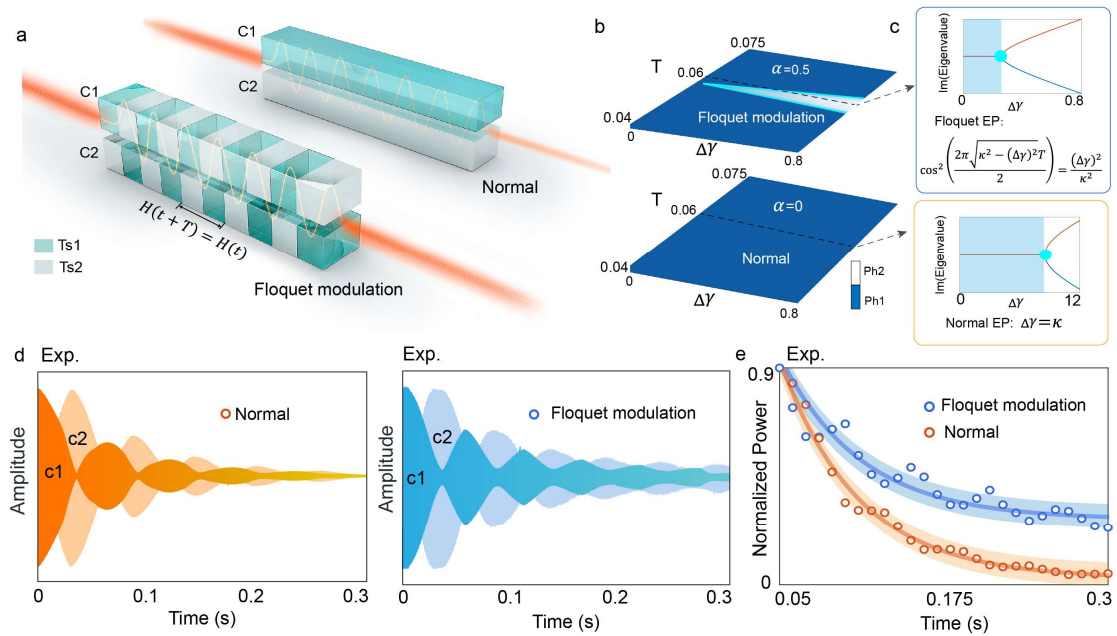

**Fig. 2 | Floquet engineering for dissipation suppression.**

**a** Schematic implementation of Floquet modulation versus static configuration, showing modulated $C1/C2$ with $Ts_1/Ts_2$ relaxation time switching. **b** Phase diagram in the parameter space $(\delta\gamma, T)$ for the static ($\alpha = 0$, bottom plane) and Floquet modulation ($\alpha =$

0.5, top plane) system under $\kappa = 8.5$. The orange star and green circle mark the experimental parameters used for the static and Floquet cases, respectively. **c** Comparison of exceptional point (EP) conditions under Floquet modulation and in the static case. **d** Experimental time-domain amplitudes (C1, C2): Static system (left, orange) shows decay, while Floquet driving (right, blue) leads to amplification and stable oscillations (parameters: $\delta\gamma = 0.4, T = 0.06$ s). **e** Experimental comparison of the normalized power decay over time. Floquet modulation (blue circles) compared to the static case (orange circles).

**Ramsey phase modulation of Bloch dynamics**

While amplitude evolution reveals population dynamics, the phase evolution encodes essential relaxation information inaccessible via intensity measurements alone. Building on the amplitude dynamics observed under continuous driving, we now turn to the phase evolution of the system using Ramsey interferometry[37,38]. This widely adopted technique probes the phase accumulated during controlled free evolution intervals and enables precise phase manipulation and measurement, offering complementary insight into the system's coherent dynamics. Fig. 3a illustrates the schematic three-step Ramsey interference protocol along with the corresponding evolution trajectories on the Bloch sphere. The protocol starts from an initial state $|0\rangle$, followed by a $\pi/2$ pulses ($H_{drive} = 2\pi\kappa\,\hat{\sigma}_x$, duration $t = 1/(8\kappa)$) to generate the superposition. Subsequently, a switched Hamiltonian ($H_{drive} = \pi\Delta\hat{\sigma}_z$) realizes a state rotation along the z-axis on the Bloch sphere. Finally, a second $\pi/2$ pulse ($H_{drive} = 2\pi\kappa\,\hat{\sigma}_x, t = 1/(8\kappa)$) projects the resulting state onto the measurement basis state $|0\rangle$ and $|1\rangle$. As a result, the phase accumulated during the precisely controlled evolution time $\tau$ manifests as characteristic periodic interference patterns—Ramsey fringes—whose period and contrast depend sensitively on the detuning $\Delta$.

To validate the effectiveness of our phase manipulation protocol, we target the observation of high-visibility Ramsey fringes and their precise dependence on the detuning $\Delta$ and evolution time $\tau$. Fig. 3b presents both theoretical predictions and experimental measurements of the $|0\rangle$ state population as functions of $\tau$ and $\Delta$, showing excellent agreement and confirming the fidelity of our coherent phase control. Furthermore, as illustrated in Fig. 3c, Bloch sphere trajectories offer an intuitive geometric interface for real-time phase manipulation:

by tuning $\tau$ and $\Delta$, the transient wavefunction can be dynamically steered along controllable phase paths. This geometric representation enables in situ visualization of phase accumulation, directly linking parametric control to interference fringe visibility through active trajectory engineering. To quantify the effects of dissipation on phase accumulation, we analyze the Ramsey interference fringe envelope by fitting the amplitude decay of the $|0\rangle$ state population as a function of evolution time. The signal follows the standard expression: $P_{|0\rangle}(\tau) = \frac{1}{2}\left[1 - e^{-\frac{\tau}{T_2^*}}\cos(\tau)\right]$, where $T_2^*$ denotes the characteristic phase relaxation time[22]. From the experimental data (Fig. 3d), we extract a relaxation time of $T_2^*=0.19$ s. These results establish a robust experimental and theoretical foundation for implementing more advanced composite pulse schemes, enabling engineered state evolution trajectories on the Bloch sphere.

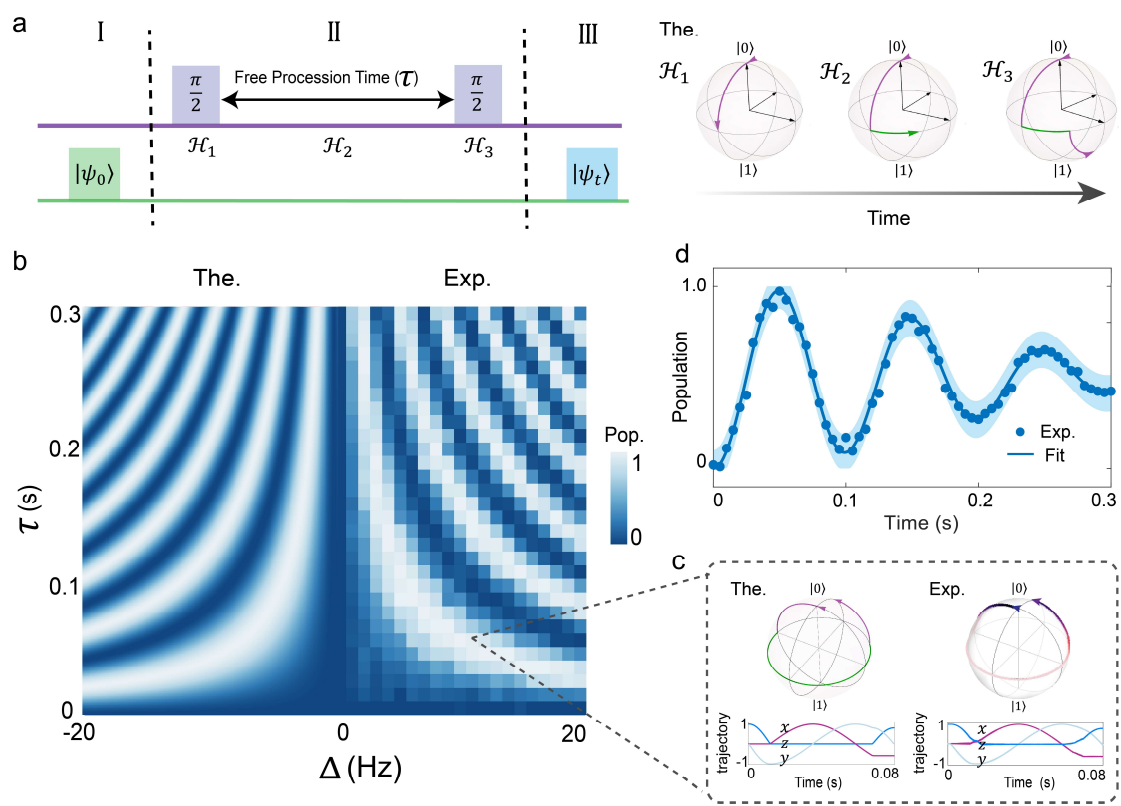

**Fig. 3| Ramsey interference in the classical electrical-acoustic two-level system.**

**a** Pulse sequence and corresponding Bloch-sphere representations illustrating the transient evolution. The system is initially prepared in mode $|0\rangle$, then driven into a superposition state by the first $\pi/2$ pulse. During the free procession interval ($\tau$), tunable mode detuning induces measurable relative phase accumulation. A second $\pi/2$ pulse translates this phase into observable Ramsey-type interference fringes. **b** Theoretical (left) and experimental measurements (right) of the $|0\rangle$ state population as functions of detuning ($\Delta$, from 0 to 20 Hz) and evolution time ($\tau$, up to 0.3 s). **c** Bloch-sphere trajectories showing high-precision, real-time phase modulation corresponding to individual interference points, demonstrating excellent agreement between theory and experiment. **d** Measurement and exponential fit of interference amplitude decay at a detuning of 10 Hz, yielding a coherence time $T_2^*$=0.19 s.

**Composite pulse Bloch dynamics in the classical electrical-acoustic TLS**

Building on the demonstrated control of amplitude and phase dynamics through Rabi and Ramsey protocols, we now extend to more sophisticated coherent control. In particular, we integrate precise phase modulation during free evolution with multi-stage acoustic pulse sequences, using the spin echo protocol as a representative example. The spin echo protocol, originally developed in nuclear magnetic resonance (NMR)[39–43], is designed to reverse phase decoherence accumulated during free evolution through a refocusing $\pi$-pulse. Its hallmark feature is that the final state rephases to a well-defined point—ideally returning to the initial state—regardless of the specific duration of free evolution. The sequence is implemented by applying a series of tailored unitary operations to the initial state $|0\rangle$ yielding the final state:

$$|\psi(T)\rangle = U_{\frac{\pi}{2}}U(\tau)U_{\pi}U(\tau)U_{\frac{\pi}{2}}|0\rangle \tag{4}$$

Here, $U_{\frac{\pi}{2}}$ ($U_{\pi}$) denotes the $\frac{\pi}{2}$ ($\pi$) pulse, and $U(\tau)$ denotes the free evolution operator for a duration $\tau$ under detuning $\Delta = 10$ Hz, as illustrated in Fig. 4a. The protocol consists of five precisely timed, millisecond-scale control intervals: $[0, t_1]$ $[t_2, t_3]$ and $[t_4, t_5]$ involve driven evolution under $H_{drive} = 2\pi\kappa\,\hat{\sigma}_x$ (corresponding to $\frac{\pi}{2}, \pi, \frac{\pi}{2}$ pulses), while $[t_1, t_2]$ and $[t_3, t_4]$ corresponding to free evolution under $H_{drive} = \pi\Delta\,\hat{\sigma}_z$. A detailed account of the wavefunction rotation and phase accumulation induced by the spin-echo pulse driving is provided in Supplementary S4. This time-reversal-like mechanism enables high-fidelity recovery of coherence and has wide applications in quantum control, NMR spectroscopy, and decoherence mitigation. As shown in Fig. 4b, to demonstrate Bloch sphere-based dynamic control, the normalized population dynamics under varying free evolution (S1-S3) times and total durations reveal a pronounced acoustic rephasing behavior. The system consistently achieves precise spatiotemporal focusing of the final acoustic field at the predetermined measurement instant, independent of the specific evolution time $\tau$. Both the measured population dynamics and the corresponding Bloch-sphere trajectories exhibit a phase retrospection effect: the state vector retraces a time-reversed path, converging back to its initial state after the composite pulse sequence.

In contrast, as illustrated in Figs. 4d and 4e, when the final pulse in the sequence is replaced by a $\frac{3\pi}{2}$ pulse, $|\psi(T)\rangle$ no longer returns to the initial ground state $|0\rangle$. Instead, the system exhibits a robust stabilization into the excited state $|1\rangle$ for any chosen free evolution interval $\tau$. The corresponding Bloch-sphere trajectories shown in Fig. 4e further highlight this behavior. Both theoretical predictions and experimental observations clearly indicate that the state vector evolves along distinctly different paths compared to the conventional spin echo sequence, ultimately converging onto the excited state. These results demonstrate the platform's ability to realize programmable, phase-sensitive state steering using complex acoustic pulse sequences. The strong agreement between experiment and theory highlights the utility of transient Bloch-sphere dynamics as a tool for real-time visualization and high-fidelity control in time-dependent acoustic-field engineering.

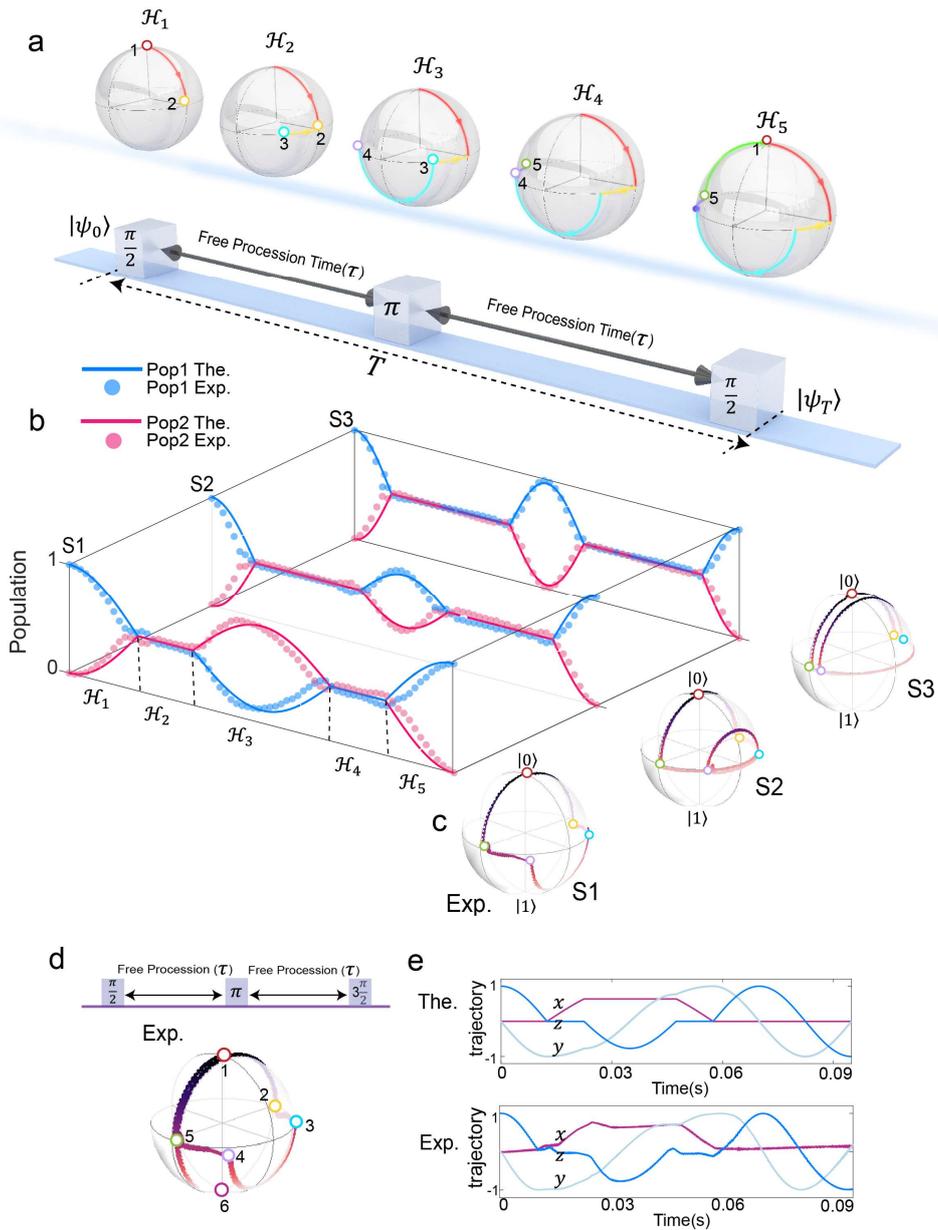

**Fig. 4 | Spin-echo dynamics. a** Schematic of acoustic spin echo sequence with corresponding Bloch-sphere state evolution. **b** Normalized population dynamics demonstrating precise acoustic control under varying free precession times of 0.01 s, 0.03 s, and 0.04 s ($\Delta = 10$ Hz, $\kappa = 10$), corresponding to S1–S3, respectively. **c** The state evolution under composite pulse driving at different free precession times (S1–S3) is intuitively visualized through Bloch-sphere trajectories. **d** Sequence with final pulse of $3\pi/2$ pulse, resulting in stable rephasing to state $|1\rangle$. **e** Corresponding theoretical (top) and experimental (bottom) trajectories of the state components, illustrating stable convergence to the state $|1\rangle$.

**Discussion**

In summary, we have experimentally demonstrated a classical acoustic two-level system capable of realizing complete Bloch sphere dynamics through programmable spatiotemporal modulation. By implementing quantum-inspired protocols including Rabi oscillations, Ramsey interference, Floquet modulation, and spin echo sequences, we achieve high-fidelity control over both amplitude and phase evolution, accompanied by real-time geometric visualization. This capability to precisely engineer transient trajectories, mitigate decoherence, and manipulate states via composite pulse sequences underscores the versatility of our platform for coherent acoustic field control. Remarkably, the programmable synthetic fields demonstrated here serve as classical analogues of electrons evolving under magnetic gauge fields. Looking forward, extending our methods to emulate dynamics of more complex particles, such as those governed by non-Abelian gauge fields, represents an exciting frontier. Such explorations would deepen the connection between classical wave systems and fundamental gauge theories, potentially enabling acoustic analogues of topological quantum phenomena and non-Abelian dynamics. Thus, our approach not only provides a robust foundation for wave-based quantum simulations and practical acoustic logic devices but also establishes a powerful, ambient-conditions platform for future interdisciplinary explorations bridging acoustics, condensed matter physics, and quantum field theory.

**Method**

**Experiment setup**

To achieve precise, programmable control of acoustic states on the Bloch sphere, we designed an electro-acoustic platform composed of two precision-machined stainless-steel cavities (C1 and C2, each 55 mm × 55 mm × 87 mm) operating at their fundamental resonance frequencies. Each cavity incorporates voltage-controlled amplifiers (VCA821) and digital programmable phase shifters (MCP4110), forming active feedback loops that precisely compensate undesired phase shifts and allow real-time control of the system's loss parameters. Additionally, the resonance frequency of cavity C1 is tuned dynamically through a programmable phase-shift circuit embedded within its gain loop. The coupling strength between cavities is finely adjusted by a synchronized coupling circuit consisting of VCAs, digital phase shifters, and optocoupler relay switches. With these innovations, our platform enables synchronized and programmable modulation of multiple system parameters on millisecond timescales. This unprecedented multi-degree-of-freedom transient control capability facilitates accurate and real-time engineering of the acoustic state trajectories, paving the way for robust classical simulations of quantum-inspired Bloch dynamics.

**Code availability**

The code used to analyze the data and generate the plots for this paper is available from the corresponding author upon request.

**Data availability**

The main data supporting the findings of this study are available within this letter and its supplementary information.

**Acknowledgements**

This work is supported by the National Key R&D Program of China (Grants No. 2023YFA1406900, No. 2022YFA1404403, No. 2021YFB3801800), the National Science Foundation of China (No. 1247043673).

**Author contributions**

Z.-G.C. conceived the idea. Z.-G.C and X.-M.Z. developed the theory. X.-M.Z., G.-C.H. and Z.-X.C. performed the experiment. All authors contributed to analyzing the data and writing the manuscript.

**Competing interests**

The authors declare no competing interests.